\documentclass[a4paper,10pt,twoside]{just}

\usepackage{multicol}
\usepackage{graphicx}
\usepackage{booktabs}
\usepackage{amssymb,bm,mathrsfs,bbm,amscd}
\usepackage[tbtags]{amsmath}
\usepackage{lastpage}
\usepackage{CJK}
\usepackage[percent]{overpic}
\usepackage{refmerge}
\usepackage{epsfig}

\begin{document}
\begin{CJK*}{GBK}{song}

\fancyhead[c]{\small  10th International Workshop on $e^+e^-$ collisions from $\phi$ to $\psi$ (PhiPsi15)}
 \fancyfoot[C]{\small PhiPsi15-\thepage}

\footnotetext[0]{Received 1 Dec. 2015}

\title{Study of the process $e^+ e^- \to K\overline{K}$ in the center-of-mass energy range 1004--1060 MeV
with the CMD-3 detector at $e^+ e^-$ VEPP-2000 collider.\thanks{Supported by National Natural Science
Foundation of China (55555555) }}

\author{%
E.A.Kozyrev$^{1,2;1)}$\email{e.a.kozyrev@inp.nsk.su},
\quad A.N.Amirkhanov$^{1,2}$,
\quad A.V.Anisenkov$^{1,2}$,
\quad V.M.Aulchenko$^{1,2}$,
\quad V.S.Banzarov$^{1}$,\\
\quad N.S.Bashtovoy$^{1}$,
\quad D.E.Berkaev$^{1}$,
\quad A.E.Bondar$^{1,2}$,
\quad A.V.Bragin$^{1}$,
\quad S.I.Eidelman$^{1,2}$,\\
\quad D.A.Epifanov$^{1}$,
\quad L.B.Epshteyn$^{1,2,3}$,
\quad A.L.Erofeev$^{1,2}$,
\quad G.V.Fedotovich$^{1,2}$,
\quad S.E.Gayazov$^{1,2}$,\\
\quad A.A.Grebenuk$^{1,2}$,
\quad S.S.Gribanov$^{1,2}$,
\quad D.N.Grigoriev$^{1,4}$,
\quad F.V.Ignatov$^{1}$,
\quad V.L.Ivanov$^{1,2}$,\\
\quad S.V.Karpov$^{1}$,
\quad A.S.Kasaev$^{1}$,
\quad V.F.Kazanin$^{1,2}$,
\quad A.N.Kirpotin$^{1}$,
\quad A.A.Korobov$^{1,2}$,\\
\quad O.A.Kovalenko$^{1,2}$,
\quad A.N.Kozyrev$^{1,2}$,
\quad I.A. Koop $^{1}$,
\quad P.P.Krokovny$^{1,2}$,
\quad A.E.Kuzmenko$^{1,2}$,\\
\quad A.S.Kuzmin$^{1,2}$,
\quad I.B.Logashenko$^{1,2}$,
\quad P.A.Lukin$^{1,2}$,
\quad K.Yu.Mikhailov$^{1,2}$,
\quad V.S.Okhapkin$^{1}$,\\
\quad A.V.Otboev$^{1}$,
\quad Yu.N.Pestov$^{1}$,
\quad A.S.Popov$^{1,2}$,
\quad G.P.Razuvaev$^{1,2}$,
\quad A.A.Ruban$^{1}$,\\
\quad N.M.Ryskulov$^{1}$,
\quad A.E.Ryzhenenkov$^{1,2}$,
\quad A.I.Senchenko$^{1}$,
\quad V.E.Shebalin$^{1}$,
\quad D.N.Shemyakin$^{1,2}$,\\
\quad B.A.Shwartz$^{1,2}$,
\quad D.B.Shwartz$^{1,2}$,
\quad A.L.Sibidanov$^{4}$,
\quad P.Yu.Shatunov$^{1}$,
\quad Yu.M.Shatunov$^{1}$,\\
\quad E.P.Solodov$^{1,2}$,
\quad V.M.Titov$^{1}$,
\quad A.A.Talyshev$^{1,2}$,
\quad A.I.Vorobiov$^{1}$,
\quad Yu.V.Yudin$^{1,2}$
}
\maketitle

\address{%
$^1$ Budker Institute of Nuclear Physics, Novosibirsk, 630090, Russia\\
$^2$ Novosibirsk State University, Novosibirsk, 630090, Russia\\
$^3$ Novosibirsk State Technical University, Novosibirsk, 630092, Russia\\
$^4$ University of Sydney, School of Physics, Falkiner High Energy Physics Department, Australia
}

\begin{abstract}
\hspace*{\parindent}
The $e^+ e^- \to K^0_{S}K^0_{L}$ and $e^+ e^- \to K^{-}K^{+}$ cross sections have been measured in the center-of-mass energy range 1004--1060 MeV for 25 energy points 
with about 2$\div$3\% systematic uncertainties. The analysis is based on 5.5 pb$^{-1}$ of integrated luminosity
collected with the CMD-3 detector at the VEPP-2000 $e^+ e^-$ collider. The measured cross section is approximated 
according to Vector Meson Dominance model as a sum of $\phi, \omega, \rho$-like amplitudes and their excitations, and $\phi(1020)$ meson parameters have been obtained.

\end{abstract}


\begin{multicols}{2} 
\section{Introduction}
\label{Introd}
Investigation of $e^+ e^-$ annihilation into hadrons at low energies provides
unique information about interaction of light quarks. Precise measurement of the 
$e^+ e^- \to K\overline{K}$  cross section allows to study properties of the light vector mesons with J$^{PC} = 1^{--}$,  and is required for the precise calculation of strong 
interaction contributions to (g-2)$_\mu$ and $\alpha$(M$_Z$) values~\cite{Davier_g_2}. 
A significant deviation of coupling constants ratio $\frac{g_{\phi \to K^{+}K^{-}}}{g_{\phi \to K_{S}K_{L}}}$ from a theoretical prediction requires new comprehensive measurement of the cross sections~\cite{Bramon}.
  
The most precise previous study of the process has been performed by the CMD-2~\cite{cmdn,cmdc}, SND~\cite{sndn} and BaBar~\cite{babarn,babarc} detectors. 
In this paper we present new measurement of the $e^+ e^- \to K^0_{S}K^0_{L}$ and $e^+ e^- \to K^{-}K^{+}$ cross section, characterized by statistical advantage and performed in the center-of-mass energy $E_{c.m.}$ range 1004--1060 MeV at 25 energy points. Also the paper contains the results of the cross section interpretation according to the Vector Meson Dominance (VMD) model.

\section{CMD-3 detector and data set}
\label{CMD3}

The Cryogenic Magnetic Detector (CMD-3)  is installed in one of two interaction regions of VEPP-2000 collider~\cite{vepp2000000}, and is described elsewhere~\cite{cmd3}.
The detector tracking system consists of the cylindrical drift chamber (DC) and double-layer cylindrical multi-wire proportional Z-chamber, both used for a trigger, and both are installed inside thin (0.2 $X_{0}$)
superconducting solenoid with 1.3 T field. DC contains 1218 hexagonal cells and allows to measure charged particle momentum with 1.5-4.5$\%$ accuracy in the 100-1000 MeV/c range,
and provides the measurement of the polar ($\theta$) and azimuth ($\phi$) angles with 20 mrad and 3.5-8.0 mrad accuracy, respectively.
An amplitude information from the DC wires is used to measure the ionization losses $dE/dx$ of charged particles with $ {\sigma}_{dE/dx}\approx$11-14\% accuracy for minimum ionization particles (m.i.p.).
A barrel liquid xenon (LXe) with 5.4 $X_{0}$ and CsI crystal with 8.1 $X_{0}$ electromagnetic calorimeters are placed outside the solenoid.
The BGO crystals with 13.4 $X_{0}$ are used as the end-cap calorimeters. Return yoke of the detector is surrounded by the scintillation counters, which are required for cosmic events veto.

To study a detector response to investigated processes and 
obtain a detection efficiency, we have developed a Monte Carlo (MC) simulation of our detector based on GEANT4~\cite{GEANT4} package, and all simulated events pass all
our reconstruction and selection procedures. The MC simulation includes photon jet radiation by initial electron or positron, calculated according to Ref.~\cite{PJGen_sibid}.

The analysis is based on 5.5 pb$^{-1}$ of integrated luminosity,
collected in two scans of $\phi(1020)$ resonance region at 25 energy points in the   
$E_{c.m.}$=1004--1060 MeV range.

The beam energy $E_{beam}$ has been monitored by using the Back-Scattering-Laser-Light system~\cite{compton} which determines $E_{c.m.}$ at each energy point with about 0.06 MeV accuracy.  
\section{$e^+ e^- \to K\overline{K}$ event selection}
\begin{figure*}[hbtp]
	\begin{center}
		\includegraphics[width=65mm]{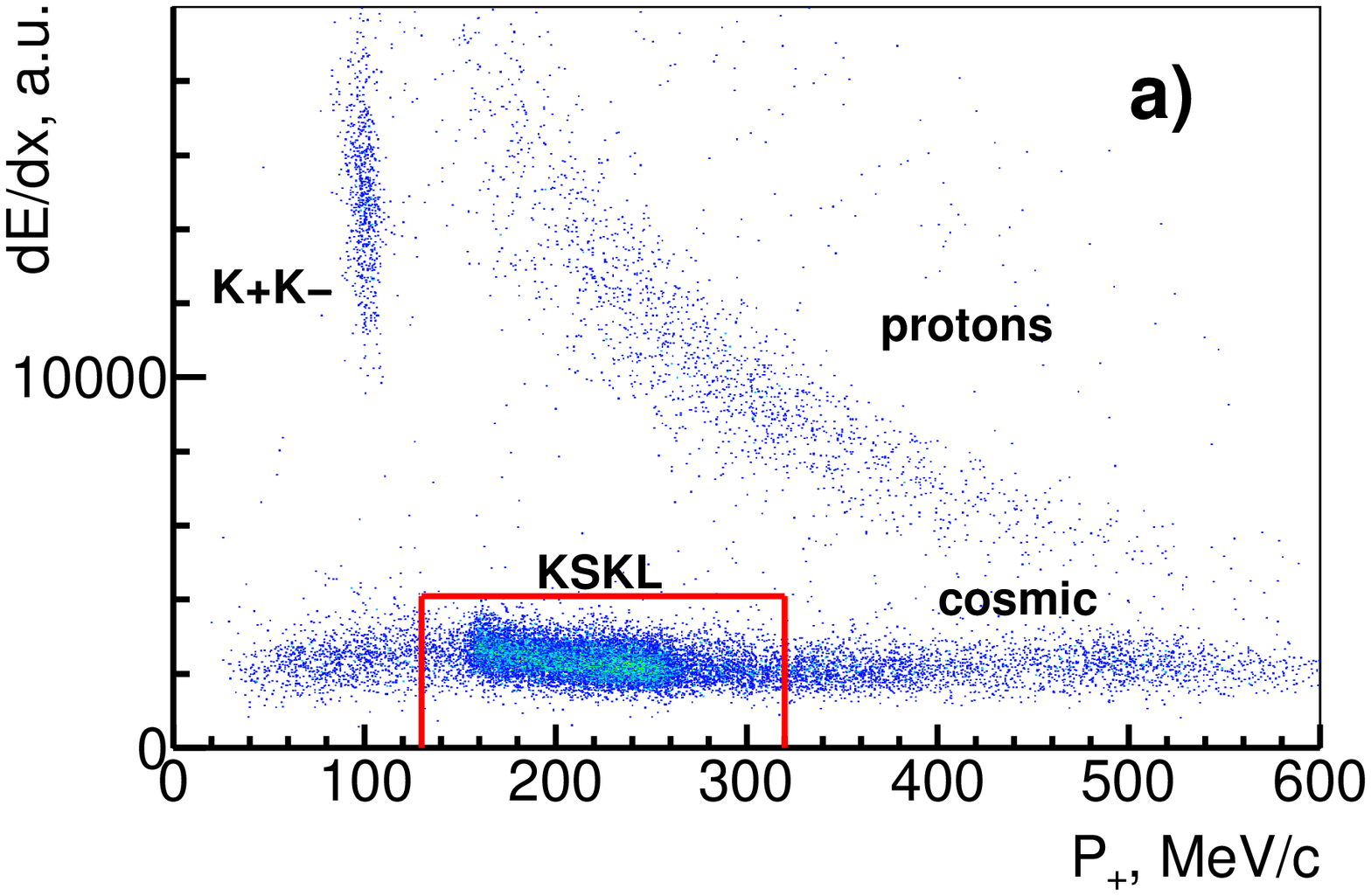}
                \includegraphics[width=65mm]{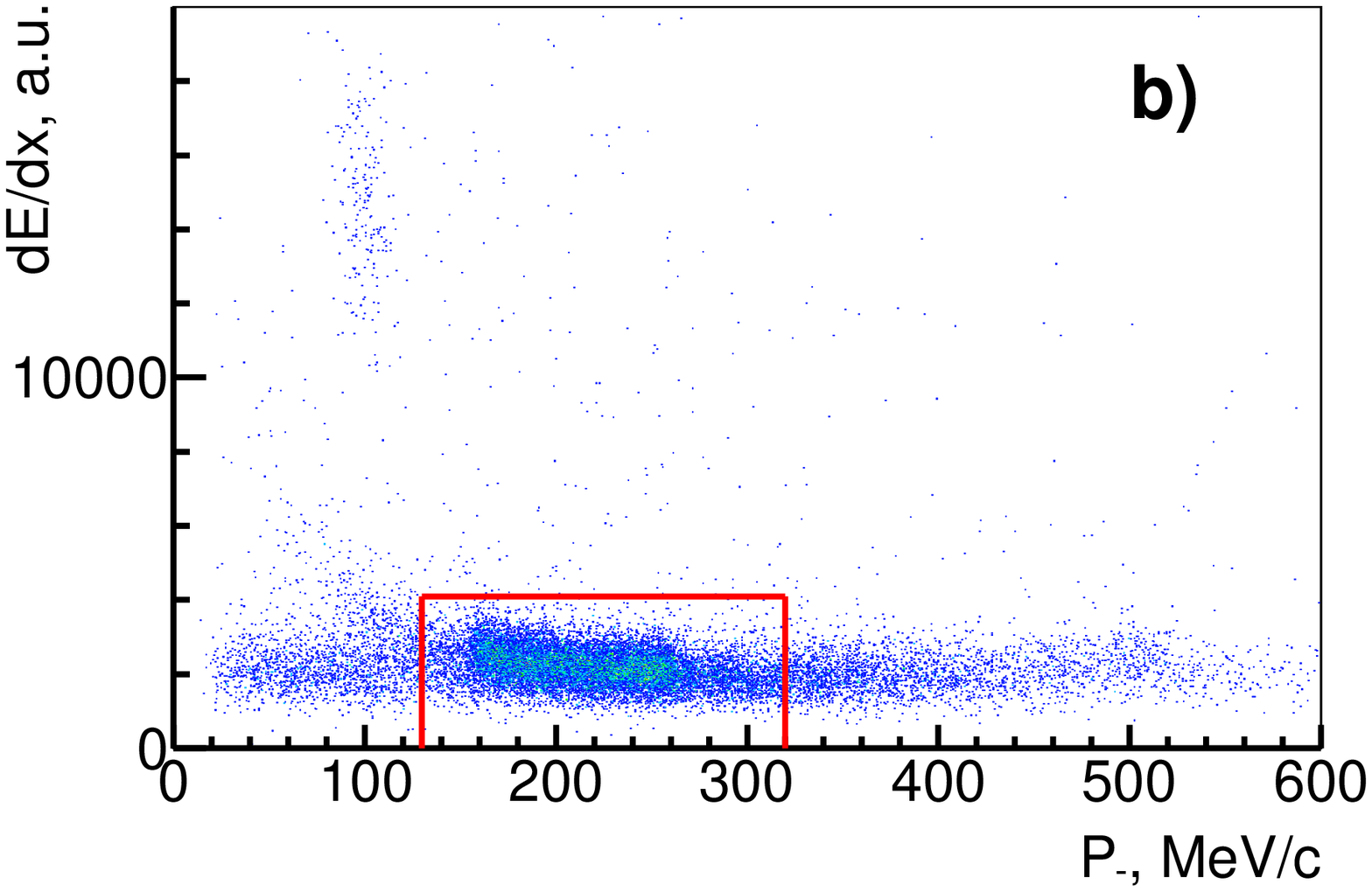}
		\figcaption{The ionization losses vs momentum  for positive (a) and negative (b) tracks for data at $E_{beam} = 505$ MeV. Lines show selections of pions from the $K^0_{S}$ decay. 
		\label{dedxp}}
	\end{center}
\end{figure*}

At energies under study $K^0_{S}$-meson can be produced only simultaneously with $K^0_{L}$-meson.
So, signal identification of the process $e^+ e^- \to K^0_{S}K^0_{L}$ is based on the detection of two pions from the $K^0_{S}\to \pi^+\pi^-$ decay. 
The following requirements are applied to the events with found $K_S^0$ candidate:
\begin{trivlist}
\item $\bullet$ The longitudinal and transverse distances of the vertex position are required to have $|Z_{K_S^0}| < 10$ cm and $|\rho_{K_S^0}| < 6$ cm, respectively;
\item $\bullet$ Each track has momentum 130 $< P_{\pi^{\pm}} <$ 320 MeV/c corresponding to the kinematically allowed region for pions from the $K_S^0$ decay;
\item $\bullet$ Each track has the ionization losses  $dE/dx_{\pi^{\pm}} < dE/dx_{m.i.p} + 3 \times \sigma_{dE/dx_{m.i.p.}}$ to reject charged kaons and background protons.
The last two requirements are illustrated in Fig.~\ref{dedxp} by lines for all detected tracks at 
the energy point $E_{beam} = 505$ MeV;
\item $\bullet$ The total reconstructed momentum of the $K^0_{S}$ candidate, $P_{K_{S}^0}$, is required to be within five standard deviations from the nominal momentum at each energy point;
\end{trivlist}
 
\begin{figure*}[hbtp]
	\begin{center}
		\includegraphics[width=65mm]{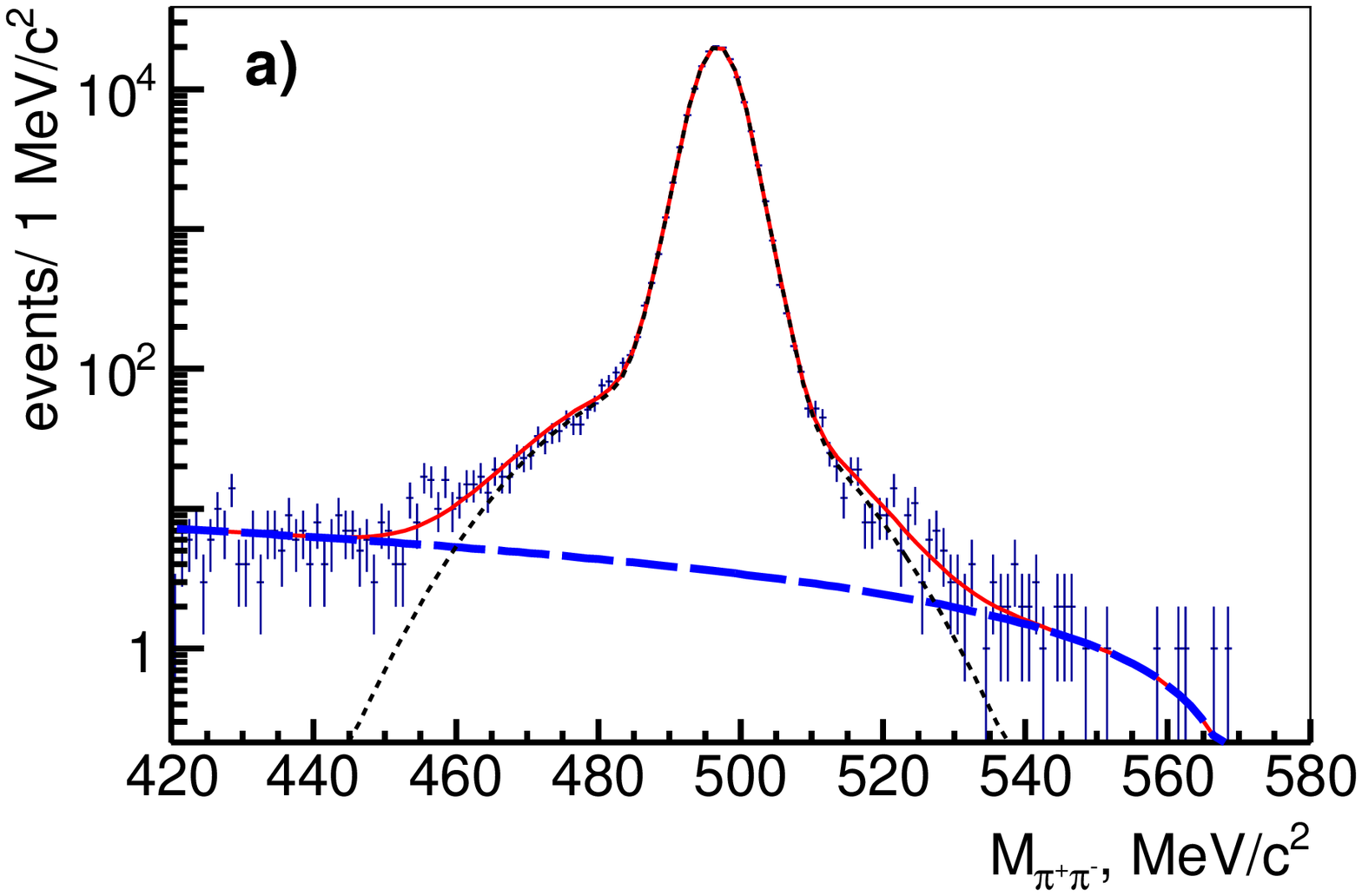}
		\includegraphics[width=65mm]{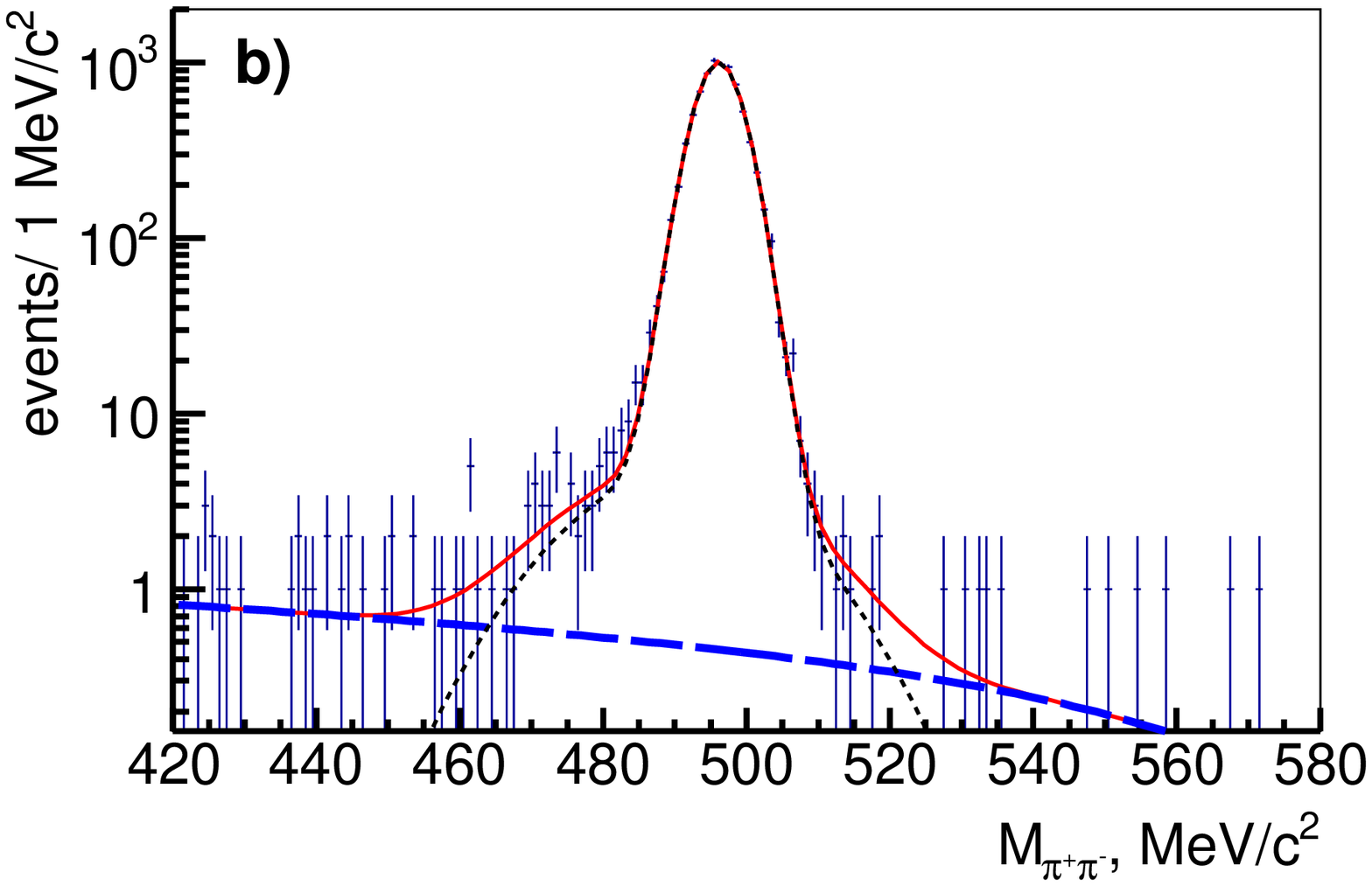}
		\figcaption{ The approximation of the invariant mass (solid line) of two pions at the beam energy  505 MeV for simulation (a) and data (b). The short dotted line corresponds to a signal profile, the long dotted line is for the background.
		\label{2pimass}}
	\end{center}
\end{figure*}

We determine number of signal events for data and simulation by approximation of two pion invariant mass, shown in the Fig.~\ref{2pimass},
by a sum of  signal and background profiles. The signal shape is described by the sum of three Gaussian functions with parameters fixed from the simulation, but with additional Gaussian smearing to account for the detector response. A background,
predominantly caused by collider processes $e^+e^-\to\pi^+\pi^-2\pi^{0}, 4\pi^{\pm}, 3\pi, K^+ K^-$ and cosmic muons, is described by second order polynomial function, and is presented in both data and MC-simulation. The background
in simulation corresponds to tails of signal with wrong reconstructed parameters of pions. By varying shapes of the functions used, we estimate uncertainty in number of extracted signal events not more than 1.1\%.

\begin{figure*}[hbtp]
\begin{minipage}[]{0.47\textwidth}
  \includegraphics[width=0.98\textwidth]{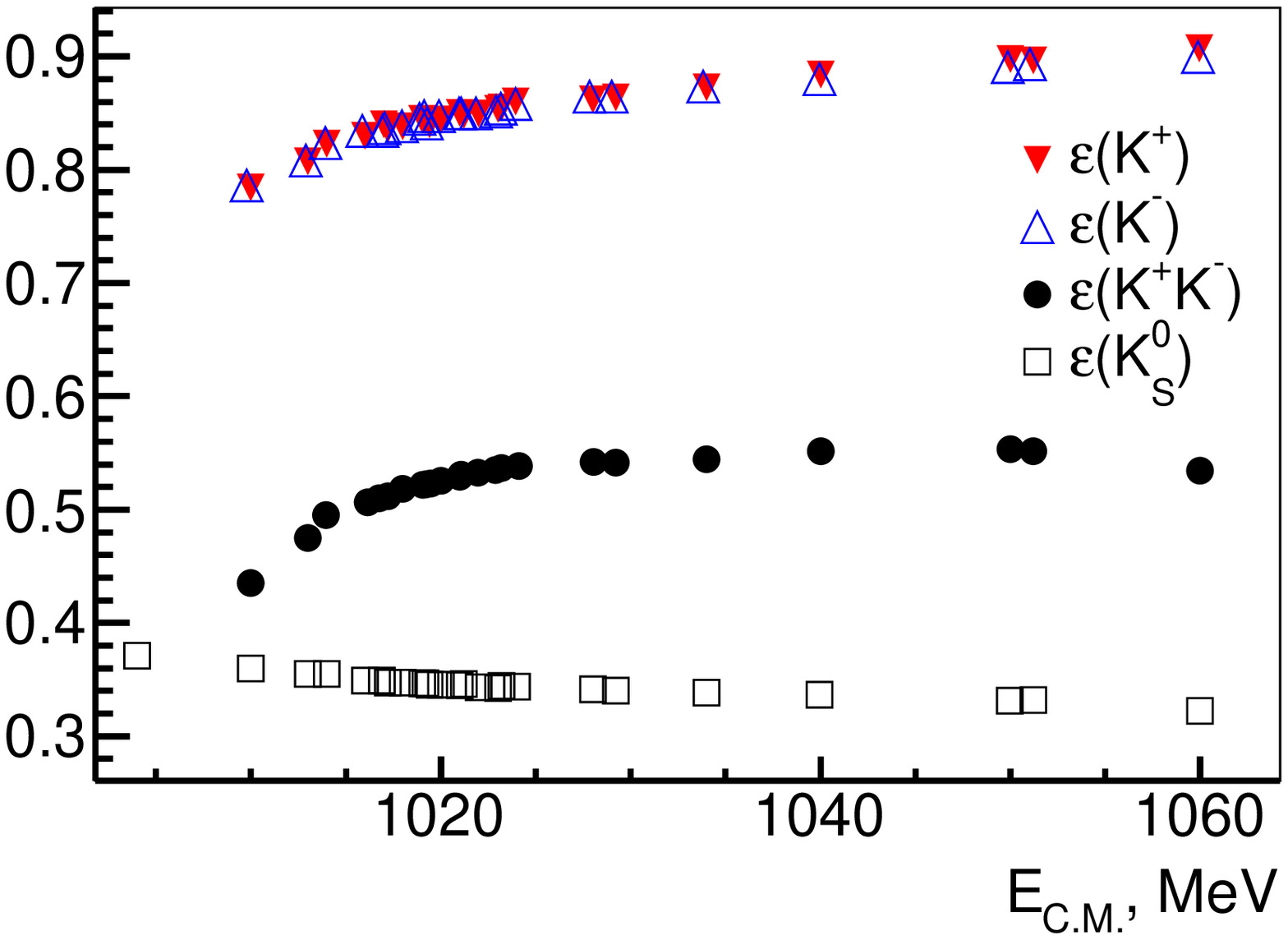}
  \figcaption{The detection efficiency vs energy from simulation;
The efficiencies of single tracks of charged kaons ($\varepsilon(K^+),~\varepsilon(K^-)$);
The efficiency of both kaons detection ($\varepsilon(K^+K^-)$) - circles;
The efficiency of $K_{S}$-meson ($\varepsilon(K^0_{S})$) - squares.
  }
  \label{efficiencypict}
\end{minipage}
\hfill
\begin{minipage}[]{0.47\textwidth}
		\includegraphics[width=0.98\textwidth]{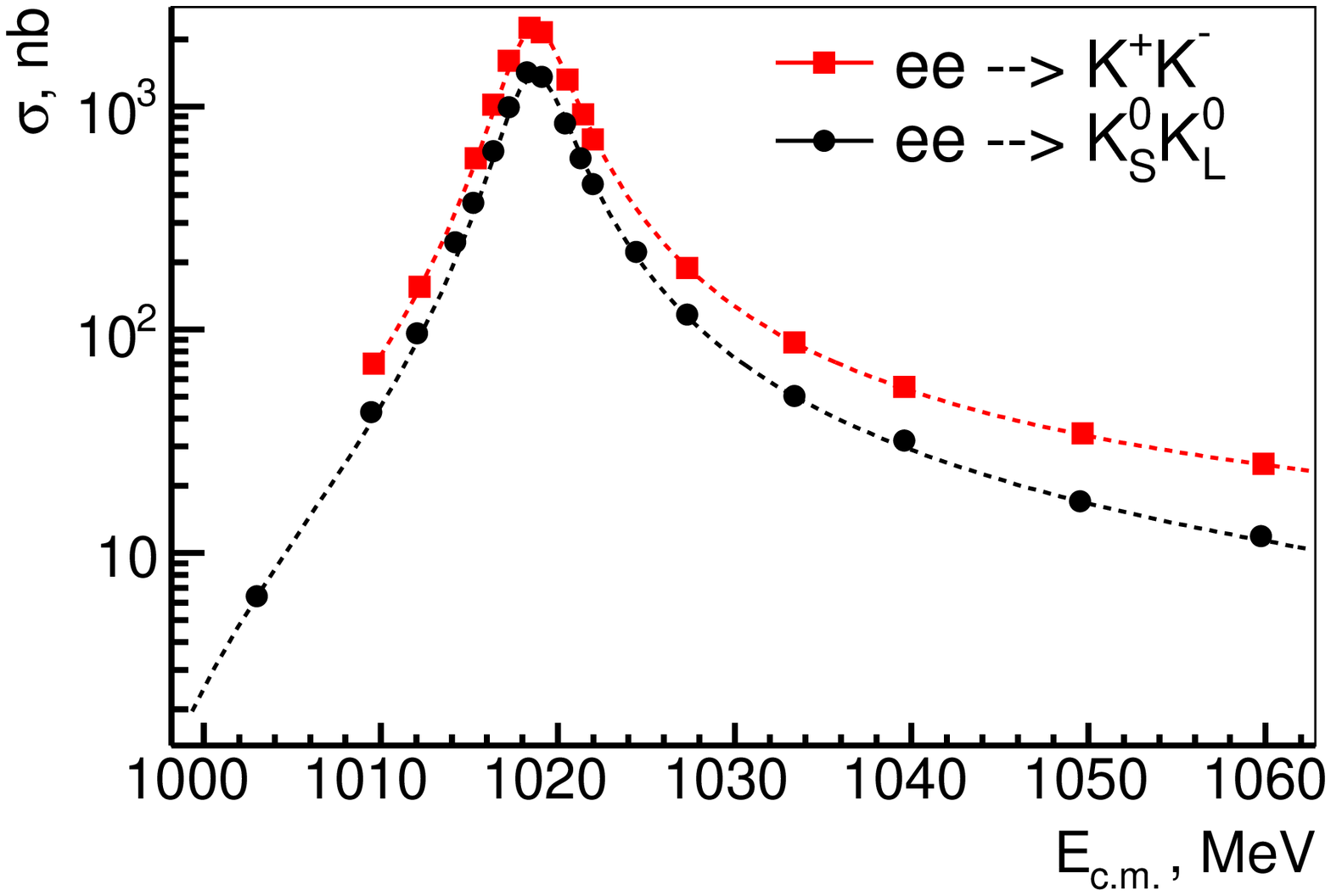}
		\figcaption{\label{crosspict} Preliminary results of measurement of the $e^+e^- \to K^+K^-$ and $e^+e^- \to K^0_{S}K^0_{L}$ cross sections near the $\phi$(1020) peak at CMD-3.}	
\end{minipage}
\end{figure*}

The detection efficiency $\epsilon(K^0_{S}K^0_{L})$ is obtained by dividing number of MC simulated events after reconstruction and selections described above, to total number of generated $K^0_{S}K^0_{L}$ pairs. Figure~\ref{efficiencypict} shows the obtained detection efficiency (squares) vs c.m. energy. 
The energy behavior as well as the absolute value ($\approx 35\%$) is predominantly due to pions polar angle selection criterion. 

%
%

The detection of the charged mode ($e^+ e^- \to K^{-}K^{+}$) is based on the search of two central collinear tracks of kaons in DC with defined momentums approximately equaled to
$\sqrt{E_{c.m.}^2/4-m_{K^+}^2}$ with accuracy of detector resolution. Additional selection exploits that kaon track has ionization losses significantly larger than the ones of m.i.p. due to relatively small velocity of kaons under study $\beta$ = 0.2-0.4 (see the Fig.~\ref{dedxp}).  
The level of remaining background is less than 0.5 \%.

The collinear configuration of the process allows to test MC simulation by the determination of the efficiencies of each kaons in data as well as in MC.
The experimental efficiencies of single positive and negative tracks ($\varepsilon(K^+),~\varepsilon(K^-)$) are shown by triangles in the Fig.~\ref{efficiencypict} and increases from 78\% to 90\% across the
energy region under study. The deviation of efficiencies of single tracks in data from MC is less then 1\%.
Circles in the figure corresponds to total simulated detection efficiency ($\varepsilon(K^+K^-$)) of $K^+K^-$ final state, that is constituted by geometrical efficiency due to polar angle selection ($\approx 73 \%$) as well as by the values of $\varepsilon(K^+),~\varepsilon(K^-)$.

\section{Cross section of $e^+ e^- \to K \overline{K}$ and systematic uncertainties}
The experimental Born cross section of the $e^+ e^- \to K \overline{K}$ process has been calculated for each energy point according to the expression: 
\begin{eqnarray}  
    \sigma^{born} &=& \frac{N_{exp}}{\epsilon_{reg}\epsilon_{trig} L (1 + \delta_{rad})}\delta^{en.disper.},  
\end{eqnarray}
where   
$\epsilon_{reg}$ is a detection efficiency, 
$\epsilon_{trig}$ is a trigger efficiency,
$L$ is the integrated luminosity, 
$1+\delta_{rad}$ is a radiative correction, and  
$\delta^{en.disper.}$ represents a correction due to the energy dispersion of the electron-positron beams.

The trigger efficiency is studied using responses of two independent triggers, charged and neutral, for selected signal events, and is found to be close to unity $\epsilon_{trigg} = 0.998 \pm 0.001$.
The integrated luminosity $L$ is determined by the processes $e^+e^- \to e^+e^-$ and $e^+e^- \to \gamma\gamma$ with about 1\%~\cite{lum} accuracy.
The initial state radiative correction $1+\delta_{rad}$ is calculated using structure function method with accuracy better than 0.3\%~\cite{radcorFadin}.
The dispersion of the electron-positron c.m. energy is about 300 keV, significant in comparison with the width of $\phi$ meson, and we introduce the correction of the cross section, which has maximum value of 1.028$\pm$0.007 for both channel in the peak of $\phi$ resonance.
The resulting cross sections are shown in Fig.~\ref{crosspict}.
\begin{table*}
\begin{center}
\tabcaption{\label{endtable_syst} Summary of systematic errors in the $e^+e^- \to K\overline{K}$ cross section measurement}
\begin{tabular}[t]{||c|c|c||}
\hline
source\ systematic error      &$e^+e^- \to K_{S}^0 K_{L}^0$  & $e^+e^- \to K^+K^-$\\
\hline
signal extraction             & 1.1    & 0.3      \\
detection efficiency       & 1.0    & 3          \\
radiative correction          & 0.3    &0.3  \\
energy dispersion correction  & 0.3    &0.3\\
trigger efficiency            & 0.1    & 0.1\\
luminosity                    & 1.0    & 1.0\\
\hline
total                         & 1.8  & 3.2    \\
\hline
\end{tabular}
\end{center}
\end{table*}

The uncertainty in $e^+ e^- \to K^0_{S}K^0_{L}$ cross section is dominated by 
signal extraction procedure used two pion mass approximation (Fig.~\ref{2pimass}).  
Moreover, MC simulation doesn't exactly reproduce all detector responses, and we perform additional study to obtain corrections for  data-MC difference in the detection efficiency. 
We observe good  data-MC agreement for the charged pion detection inefficiency ($\approx1\%$), introduce no efficiency correction, and estimate uncertainty in the detection as 0.5\%.
By variation of corresponding selection criteria we estimate uncertainty due to the data-MC difference in the angular and momentum resolutions as 0.4\%, and other selection criteria contribute 0.5\%. 

The uncertainty in $e^+ e^- \to K^+K^-$ cross section is dominated by not exact knowledge of angular acceptance of kaons.
This systematic uncertainty (3\%) are examined using Z-chamber which surrounds DC.  Unlike pions in neutral channel the charged pair of kaons have much more ionization losses and collinear configuration that leads to strong correlation in detector response to charged kaons tracks.

The systematic errors of the $e^+e^- \to K_{S}^0 K_{L}^0$ and $e^+e^- \to K^+ K^-$ cross sections measurement, discussed above, are summarized in Table~\ref{endtable_syst}, and in total estimated as 1.8\% and 3.2\% respectively.

\section{Cross section interpretation}
Our data in the studied energy range allows to obtain $\phi(1020)$ parameters with good accuracy.
We approximate the energy dependence  of the cross section according to a vector meson dominance (VMD) model as a sum of $\phi,~\omega,~\rho$-like amplitudes~\cite{Kuhn}:
\begin{eqnarray} 
\sigma_{e^+ e^- \to K \overline{K}}(s) = \frac{8 \pi \alpha}{3 s^{5/2}} p_{K}^{3} \frac{Z(s)}{Z(m_{\phi^2})} | \frac{g_{\phi \gamma} g_{\phi KK}}{D_{\phi}(s)}
\pm \frac{g_{\rho \gamma} g_{\rho KK}}{D_{\rho}(s)} 
+ \nonumber \\ 
\frac{g_{\omega \gamma} g_{\omega KK}}{D_{\omega}(s)} + A_{\phi',\rho',\omega'}|^{2},
\label{ksklxs}
\end{eqnarray}
where $s = E_{c.m.}^2$, $p_{K}$ is a kaon momentum, Z(s) $ = 1 + \frac{\pi\alpha}{2\beta}$ - the Sommerfeld-Gamov-Sakharov factor for charged kaons with velocity $\beta = \sqrt{1-4m_{K}/s}$, $D_{V}(s) = m_{V}^{2} -s - i \sqrt{s}\Gamma_{V}(s)$,
$m_{V},$ and $\Gamma_{V} $ are mass and width of major intermediate resonances: $V = \rho(770), \omega(782), \phi(1020)$. 
The sign before $\rho$-amplitude is plus for charged channel and minus for neutral one due to quark  structure of kaons and $\rho$-meson. 
The energy dependence of the decay width is expressed via sum of branching fractions and phase space energy dependence $P_{V\to f}(s)$ of all decay modes as (see~\cite{sndn,Kuhn}): 
\begin{eqnarray*}
\Gamma_{V}(s) = \Gamma_{V} \sum_{V\to f} B_{V\to f} \frac{P_{V\to f}(s)}{P_{V\to f}(m_V^2)}.
\end{eqnarray*}

The coupling constants of the intermediate vector meson $V$ with initial and final states can be presented as: 
\begin{eqnarray*}
g_{V \gamma} = \sqrt{\frac{3 m_{V}^{3} \Gamma_{Vee}}{4 \pi \alpha}};~g_{V K\overline{K}} = \sqrt{\frac{6 \pi m_{V}^{2} \Gamma_{V} B_{VK\overline{K}}}{p^{3}_{K}(m_{V})}},
\end{eqnarray*}
where $\Gamma_{Vee}$ and $B_{VK\overline{K}}$ are electronic width and decay branching fraction to pair of kaons.

In our approximation we use table values of mass, total width, and electronic width of $\rho(770)$ and $\omega(782)$:  $\Gamma_{\rho\to ee} = 7.04\pm0.06~\rm keV,~ \Gamma_{\omega\to ee} = 0.60\pm0.02~\rm keV$~\cite{PDG}.
The branching fractions of $\rho(770)$ and $\omega(782)$ to pair of kaons unknown, and 
we use the relation $g_{\omega K^0_{S}K^0_{L}} = - g_{\rho K^0_{S}K^0_{L}} = g_{\phi K^0_{S}K^0_{L}}/\sqrt{2}$, 
based on the quark model with "ideal" mixing and exact SU(3) symmetry of u-,d-,s-quarks~\cite{Kuhn}. In order to take into account possible breaking of the assumption both $g_{\rho K^0_{S}K^0_{L}}$ and $g_{\phi K^0_{S}K^0_{L}}$ are multiplied to the union constant $r_{\rho/\omega}$.  

The amplitude  $A_{\phi',\rho',\omega'}$ denotes a contribution of  excited $\omega(1420),~\rho(1450)$ and $\phi(1680)$ vector meson states to the $\phi(1020)$ mass region.
Using BaBar~\cite{babarn, babarc} data above 1.06 GeV for the $e^+e^- \to K^0_{S}K^0_{L}$ and $e^+e^- \to K^{+}K^{-}$ reactions we fix the contribution of higher energy states. 

We fit the cross sections of $e^+e^- \to K_{S}^0 K_{L}^0$ and $e^+e^- \to K^{+} K^{-}$    with float $m_{\phi},~\Gamma_{\phi}$, $\Gamma_{\phi \to e^{+}e^{-}}\times B_{\phi \to K^0_S K^0_L}$, $\Gamma_{\phi \to e^{+}e^{-}}\times B_{\phi \to K^{+} K^{-}}$ and $g_{\rho/\omega}$ parameters.
The obtained fit is shown in the Fig.~\ref{crosspict} with the following parameters, which contain statistical errors as well as systematic and model-dependent uncertainties:
\begin{eqnarray} 
\label{phipar}
m_{\phi} = 
1019.464 \pm 0.060 ~\rm  MeV/c^2  \\
\Gamma_{\phi} = 
4.247 \pm  0.015 ~\rm  MeV \nonumber \\
\Gamma_{\phi \to ee} B_{\phi \to K^0_S K^0_L} = 
 0.429 \pm 0.009 ~\rm keV \nonumber \\
\Gamma_{\phi \to ee} B_{\phi \to K^{+} K^{-}} = 
0.679 \pm 0.022~\rm keV \nonumber \\
r_{\rho/\omega} =  
0.76 \pm 0.11 \nonumber \\
g_{V \to K^{+} K^{-}}/g_{V \to K_{S}^0 K_{L}^0} = 0.995 \pm 0.035 \nonumber
\end{eqnarray}

\begin{figure*}[hbtp]
\begin{center}
   \begin{overpic}[scale=0.9]{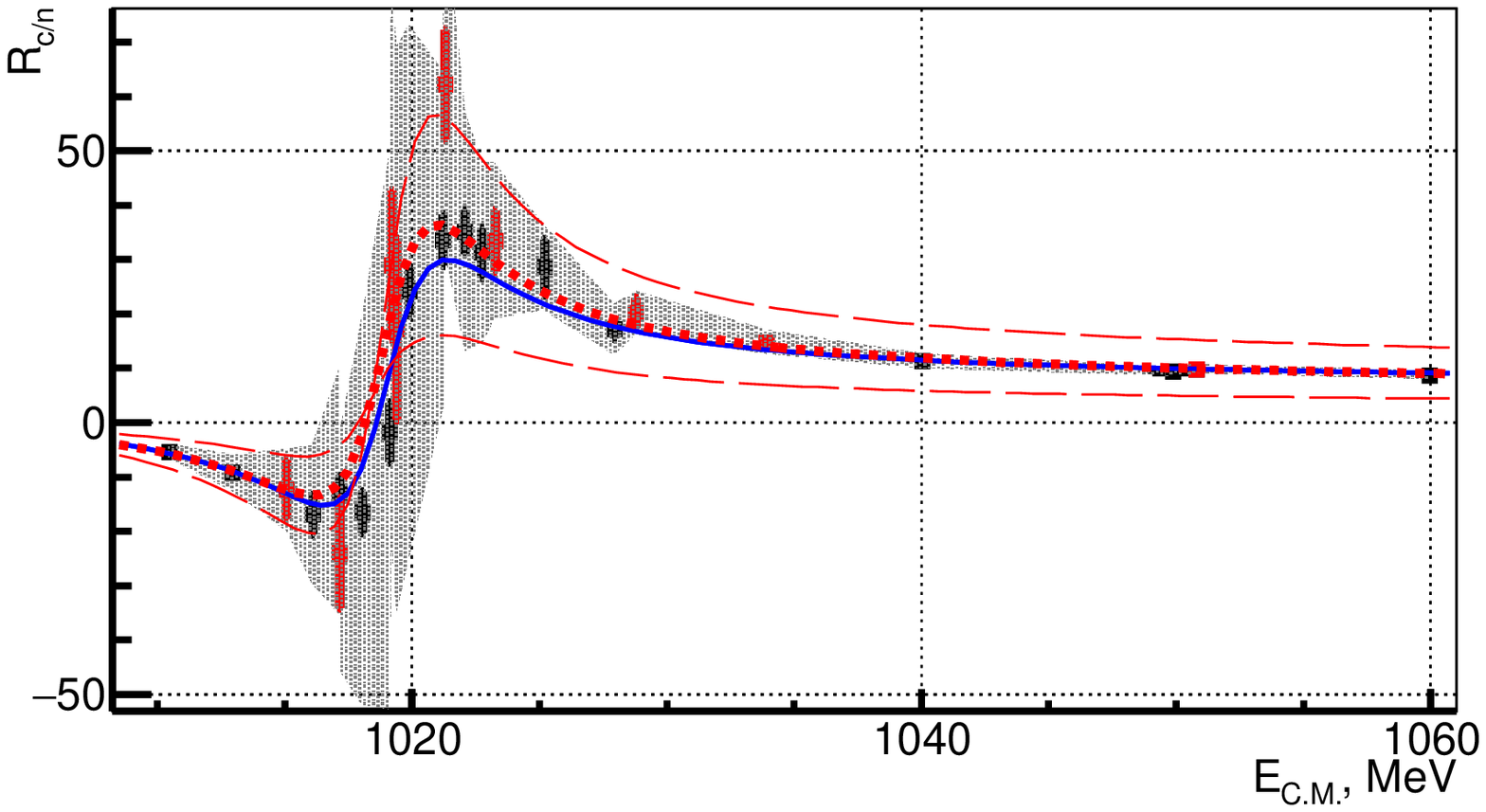}
 \put(55,34){\includegraphics[scale=0.27]{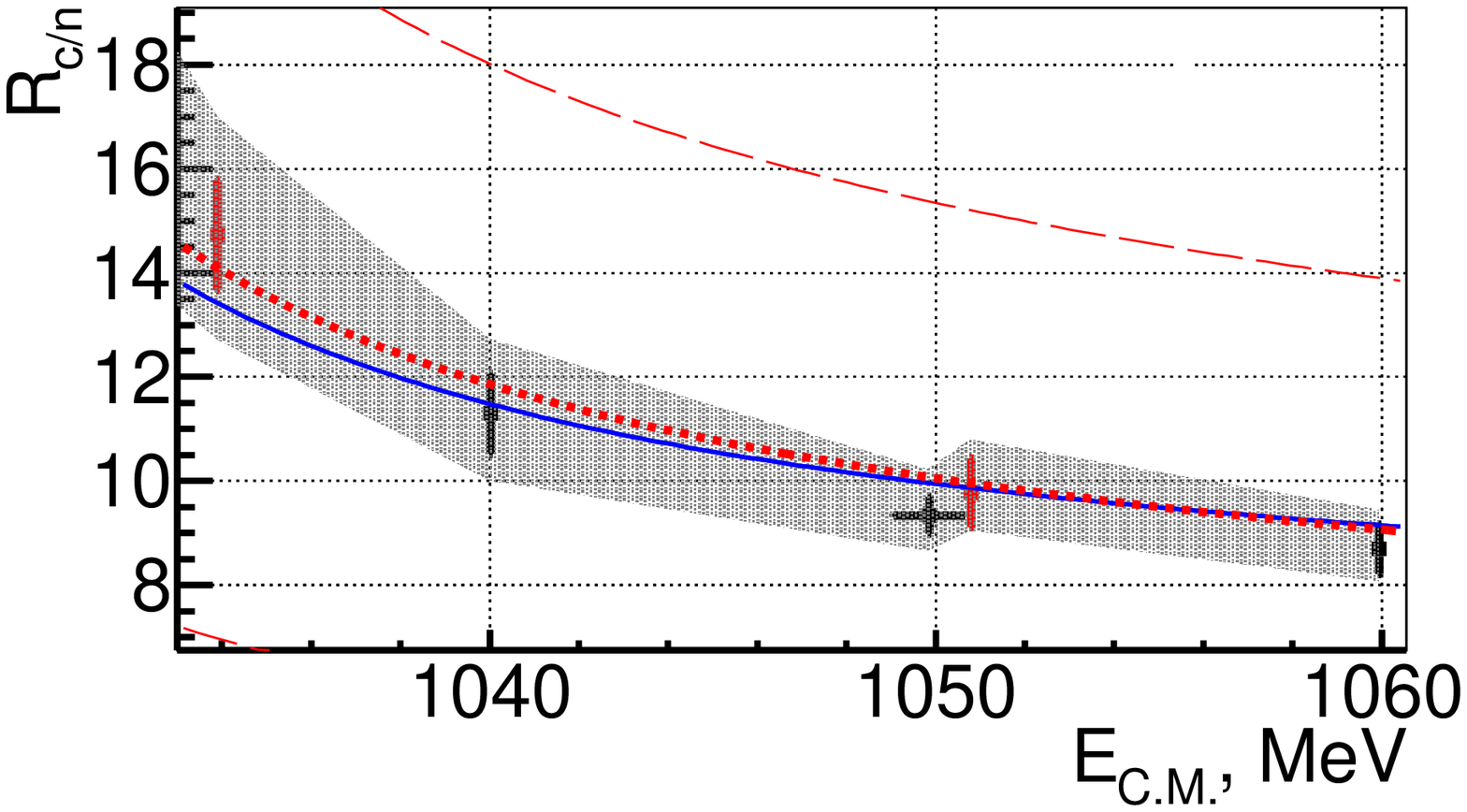}}
 \put(55,11.5){\includegraphics[scale=0.27]{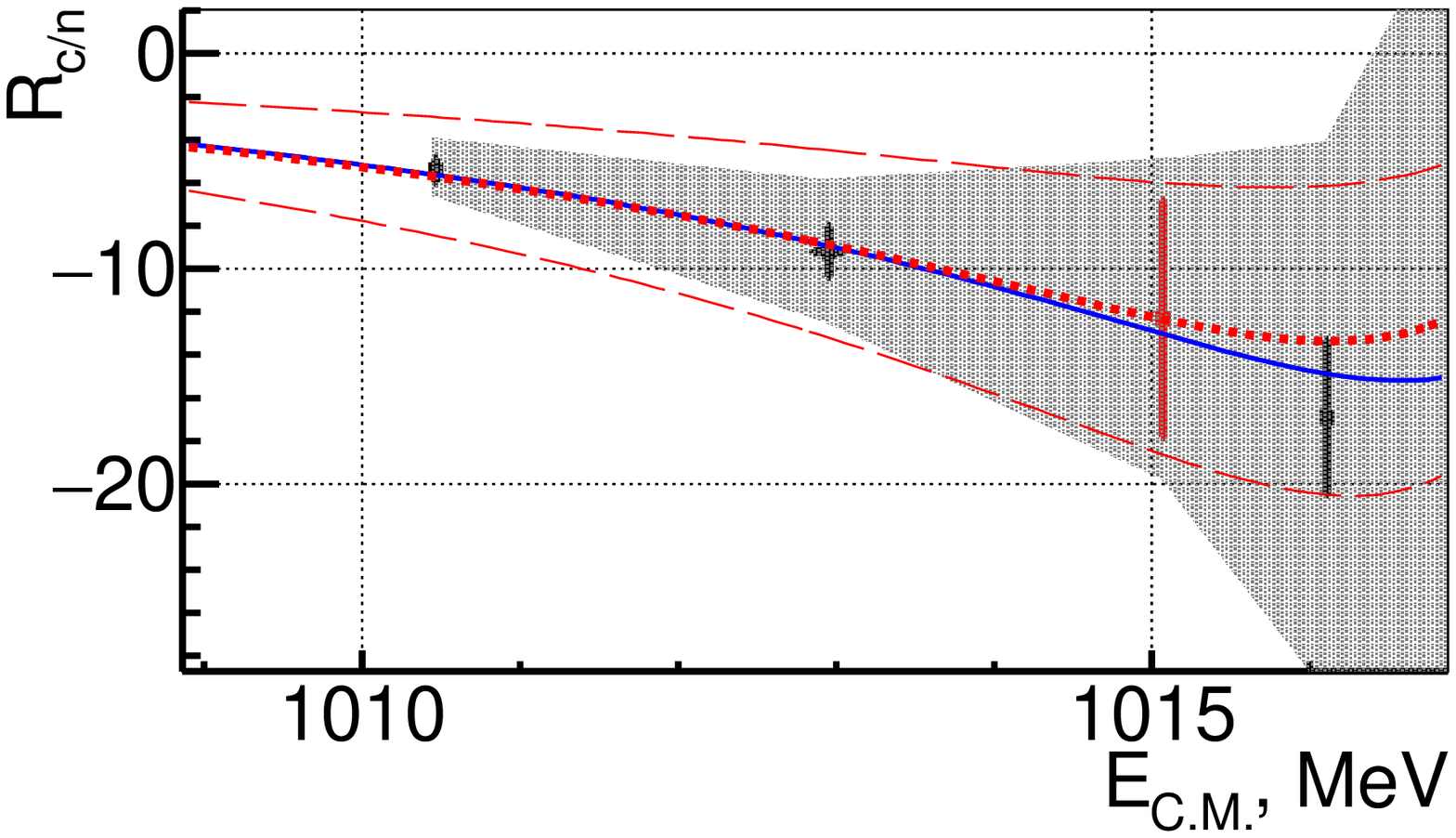}}
\end{overpic}
  \caption{ \label{cross_phi_rel}   
The difference of charged and neutral cross-sections defined as
$R_{c/n} = \sigma_{e^+e^- \to K^{+} K^{-}}\times \frac{p_{K^0}^3(s)}{p_{K^{\pm}}^3(s)} \times \frac{1}{Z(s)} - \sigma_{e^+ e^- \to K^0_{S}K^0_{L}}$.
Shaded area corresponds to systematic uncertainties in data, solid line - to fit of data,
short dotted line - to theoretical prediction  with $A_{\phi',\rho',\omega'}$ = 0,
 $g_{V \to K^{+} K^{-}}/g_{V \to K_{S}^0 K_{L}^0}$ = 1 and  $r_{\rho/\omega}$ = 1
while long dotted lines - to the same theoretical prediction with $r_{\rho/\omega}$ = 0.5 (1.5).
}
  
\end{center}
\end{figure*}

The difference of charged and neutral cross-sections for 24 energy points defined as
$R_{c/n} = \sigma_{e^+e^- \to K^{+} K^{-}}\times \frac{p_{K^0}^3(s)}{p_{K^{\pm}}^3(s)} \times \frac{1}{Z(s)} - \sigma_{e^+ e^- \to K^0_{S}K^0_{L}}$
is shown in Fig.~\ref{cross_phi_rel}.
The difference $R_{c/n}$ is predominantly caused by interference term of resonance amplitudes of $\phi$-meson and isovector $\rho$-meson. 
Shaded area corresponds to 1.8\% and 3.2\% systematic uncertainties in data for
neutral and charged channel respectively.
The result of discussed above fit is shown by solid line that leads to agreement $\chi^2 = 37$.
It should be mentioned that the case with $A_{\phi',\rho',\omega'}$ = 0 and naive theoretical prediction $g_{V \to K^{+} K^{-}}=g_{V \to K_{S}^0 K_{L}^0}$, $r_{\rho/\omega}$ = 1 also gives an adequate description of experimental $R_{c/n}$. This case is characterized by $\chi^2 = 51$ and
shown by short dotted line,
while long dotted lines correspond to the same theoretical prediction with $r_{\rho/\omega}$ = 0.5 or 1.5 and strongly differ from data.

\section{Conclusion and acknowledgements}
Using pions from the $K^0_{S}\to\pi^+\pi^-$ decay and collinear charge kaons in DC we observe 6.5$\times$10$^5$ and 1.6$\times$10$^6$ events of 
the $e^+ e^- \to K^0_{S}K^0_{L}$ and $e^+e^- \to K^{+} K^{-}$ processes respectively  in the 1004--1060 MeV c.m. energy range, and measure its cross section with 1.8$\div$3.2\% systematic error. 
Using VMD model the parameters of $\phi$-meson are preliminary measured (\ref{phipar}).
The obtained deviation of $\rho,~\omega$ amplitudes from naive theoretical prediction  
$r_{\rho/\omega} =  0.76 \pm 0.11$ allows to estimate the precision of used VMD-based phenomenological model as 25 \%. Moreover, obtained ratio $g_{V \to K^{+} K^{-}}/g_{V \to K_{S}^0 K_{L}^0} = 0.995 \pm 0.035$ demonstrates the precision of SU(2)-symmetry better than 3.5 \%. 

We thank
the VEPP-2000 personnel for the excellent machine operation.
This work is supported in part by the Russian Education and
Science Ministry (grant No. 14.610.21.0002, identification number RFMEFI61014X0002), by the Russian Foundation for Basic
Research grants RFBR 13-02-00991-a, RFBR 13-02-00215-a, RFBR
12-02-01032-a, RFBR 13-02-01134-a, RFBR 14-02-00580-a, RFBR 14-02-31275-mol-a, RFBR 14-02-00047-a, RFBR 14-02-31478-mol-a, RFBR 14-02-91332 and the DFG grant HA 1457/9-1.

\end{multicols}

\clearpage

\end{CJK*}
\end{document}